\documentclass[reprint,twocolumn,showpacs,amsmath,amssymb,superscriptaddress,aps,pra]{revtex4-1}

\usepackage{color}
\usepackage{graphicx}
\usepackage[english]{babel}
\usepackage[hidelinks]{hyperref}

\begin{document}

\title{Sideband cooling of nearly degenerate micromechanical oscillators in a multimode optomechanical system}

\author{C. F. Ockeloen-Korppi}\affiliation{Department of Physics, Aalto University, P.O. Box 15100, 00076 Aalto, Finland.}
\author{M. F. Gely}\affiliation{Kavli Institute of NanoScience, Delft University of Technology, P.O. Box 5046, 2600 GA, Delft, The Netherlands.}
\author{E. Damsk\"agg}\affiliation{Department of Physics, Aalto University, P.O. Box 15100, 00076 Aalto, Finland.}
\author{M. Jenkins}\affiliation{Kavli Institute of NanoScience, Delft University of Technology, P.O. Box 5046, 2600 GA, Delft, The Netherlands.}
\author{G. A. Steele}\email[]{g.a.steele@tudelft.nl}\affiliation{Kavli Institute of NanoScience, Delft University of Technology, P.O. Box 5046, 2600 GA, Delft, The Netherlands.}
\author{M. A. Sillanp\"a\"a}\email[]{mika.sillanpaa@aalto.fi}\affiliation{Department of Physics, Aalto University, P.O. Box 15100, 00076 Aalto, Finland.}

\date{\today}

\begin{abstract}
Multimode optomechanical systems are an emerging platform for studying fundamental aspects of matter near the quantum ground state and are useful in sensitive sensing and measurement applications. We study optomechanical cooling in a system where two nearly degenerate mechanical oscillators are coupled to a single microwave cavity. Due to an optically mediated coupling the two oscillators hybridize into a bright mode with strong optomechanical cooling rate and a dark mode nearly decoupled from the system. We find that at high coupling, sideband cooling of the dark mode is strongly suppressed. Our results are relevant to novel optomechanical systems where multiple closely-spaced modes are intrinsically present.
\end{abstract}

\maketitle

The field of cavity optomechanics~\cite{doi:10.1103/RevModPhys.86.1391} opens possibilities for studying the fundamental quantum properties of matter and is a valuable platform for precision measurements~\cite{doi:10.1103/PhysRevLett.116.061102}. Of particular interest are optomechanical systems incorporating more than one mechanical oscillator, enabling the study of multimode effects. With mechanical modes of nearly identical frequency, coherent mode mixing and mechanical energy transfer have been demonstrated, utilizing either a direct~\cite{doi:10.1103/PhysRevLett.109.037205} or an optically induced~\cite{doi:10.1038/nphoton.2010.5,doi:10.1038/ncomms1993,doi:10.1103/PhysRevLett.112.013602,doi:10.1038/nature18604,doi:10.1038/s41467-017-00968-9} coupling between the mechanical modes. They have been applied as synchronized oscillators~\cite{doi:10.1103/PhysRevLett.109.233906} and displacement sensors~\cite{doi:10.1063/1.4889804}. When the mechanical modes are well-separated in frequency, optically mediated coupling has been demonstrated using a low-finesse cavity~\cite{doi:10.1038/nphys3515} and using multi-tone parametric driving~\cite{doi:10.1088/2040-8978/18/10/104003,doi:10.1088/1367-2630/18/10/103036}. Such systems have useful applications, including recently demonstrated non-reciprocal optical devices~\cite{doi:10.1038/nphys4009,doi:10.1103/PhysRevX.7.031001,doi:10.1038/s41467-017-00447-1,doi:10.1038/ncomms13662,doi:10.1038/s41467-017-01304-x,doi:10.1038/nphoton.2016.161}, and are of fundamental interest, with recent demonstrations of quantum back-action evading position sensing~\cite{doi:10.1103/PhysRevLett.117.140401,doi:10.1038/nature22980} and entanglement of mechanical oscillators~\cite{doi:10.1038/s41586-018-0036-z,doi:10.1038/s41586-018-0038-x}. Multi-mode optomechanics is also intrinsically present in some novel optomechanical systems utilizing high mechanical overtones, such as bulk acoustic wave resonators~\cite{doi:10.1126/science.aao1511,doi:10.1103/PhysRevB.97.205443}, where neighboring modes may affect the mode of interest.

\begin{figure}
\includegraphics[width=\columnwidth]{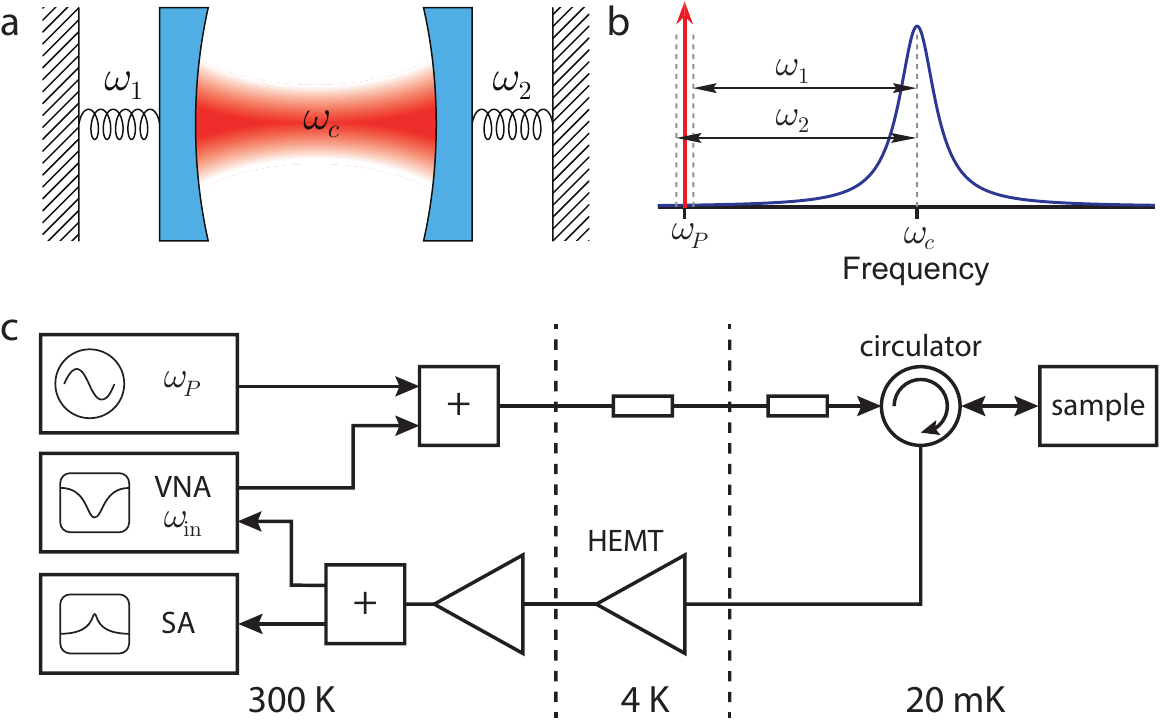}
\caption{\emph{Setup.} (a) Our multimode optomechanical system is analogous to an optical cavity with two compliant end mirrors. (b) Pumping scheme. (c) Cryogenic reflection measurement setup. The pump tone $\omega_P$ is combined with the probe tone $\omega_\text{in}$ from a vector network analyzer (VNA) and coupled to the sample. The reflected signal is amplified and recorded by either the VNA or a signal analyzer (SA).}
\label{fig:setup}
\end{figure}

The prototypical two-mode optomechanical system is a Fabry-P\'{e}rot cavity where both end mirrors are suspended on springs, as shown in Fig.~\ref{fig:setup}a. This realizes the optomechanical Hamiltonian
\begin{equation}\label{eq:hamiltonian}
H_0=\omega_c a^\dagger a+\sum_{i=1,2}\left[\omega_ib_i^\dagger b_i+ g_i a^\dagger a\left(b_i^\dagger+b_i\right)\right],
\end{equation}
where the cavity with frequency $\omega_c$ and linewidth $\kappa$ is described by the bosonic operators $a$, the mechanical modes with frequencies $\omega_i$ and linewidths $\gamma_i$ are described by $b_i$, and $g_i$ is the corresponding single-photon optomechanical coupling strength. In the presence of a coherent optical pump exciting the cavity mode by $n_P \gg 1$ quanta, and moving to a frame rotating at the pump frequency $\omega_P = \omega_c + \Delta$, the system can be linearized around the mean pump amplitude and described by the effective Hamiltonian
\begin{equation}\label{eq:linearhamiltonian}
H = -\Delta a^\dagger a + \sum_{i=1,2} \left[\omega_ib_i^\dagger b_i+ G_i\left(a^\dagger +  a\right)\left(b_i^\dagger+b_i\right)\right],
\end{equation}
where $G_i = g_i \sqrt{n_P}$ is the cavity-enhanced coupling rate. 

Sideband cooling is a standard technique in cavity optomechanics to reduce the mechanical mode temperature, giving access to the quantum ground state in a hot environment~\cite{doi:10.1038/nature10261,doi:10.1038/nature10461}. The system is driven with a strong coherent pump field, red-detuned from the cavity at frequency $\omega_P < \omega_c$. Considering only a single mechanical mode with frequency $\omega_1$, and in the resolved sideband regime $\kappa \ll \omega_1$, optimal cooling is achieved at $\Delta = -\omega_1$. Anti-Stokes scattering of pump photons off the mechanical oscillator give rise to optomechanical damping with a rate $\gamma_\text{opt} = 4 G_1^2/\kappa$, increasing the mechanical linewidth to $\gamma_\text{eff} = \gamma_1 + \gamma_\text{opt}$~\cite{doi:10.1103/RevModPhys.86.1391}.

In the presence of two mechanical modes $\omega_i$ ($i = 1,2$) coupled to a single cavity, the process of sideband cooling becomes more complex due to hybridization of the mechanical modes via the cavity. In Ref. \cite{doi:10.1088/1367-2630/10/9/095009} it was predicted that cooling of the individual mechanical modes is suppressed for $\gamma_\text{eff} > |\omega_2 - \omega_1|$, i.e., when the individual mechanical modes have significant spectral overlap and become effectively degenerate. This transition between nondegenerate and degenerate modes marks an exceptional point, which has interesting topological features such as recently demonstrated topological energy transfer~\cite{doi:10.1038/nature18604}. As was analyzed in Ref.~\cite{doi:10.1038/ncomms1993}, in this degenerate regime the (originally uncoupled) mechanical modes experience a cavity-mediated effective coupling $G_{12} = G_1G_2/\kappa$. For $\gamma_\text{eff} \ll \kappa$, the mechanical modes can still be described independent of the optical mode, by adiabatically eliminating the cavity. For the resulting two-mode degenerate mechanical system, the dissipative optomechanical coupling $G_{12}$ realizes an anti-PT symmetric Hamiltonian, a novel concept that was recently demonstrated in atomic spin systems~\cite{doi:10.1038/nphys3842}.

In the strong coupling limit $G_1^2 + G_2^2 \gg \kappa^2$, the system can be described as a combination of two optomechanical ``bright'' modes with linewidths $\approx \kappa/2$ and one all-mechanical ``dark'' mode with long lifetime $\approx \gamma_{1,2}$.  The suppression of sideband cooling in the strong coupling regime can be intuitively understood in terms of the bright and dark normal modes: while the bright modes can be effectively cooled through the optical field, the dark mode is decoupled from the cavity and not affected by the cooling. Tri-partite mixing in a near-degenerate multimode optomechanical system has been demonstrated previously~\cite{doi:10.1038/ncomms1993}, and similar mixing has been achieved in a system with clearly non-degenerate mechanical frequencies using a two-tone driving scheme ~\cite{doi:10.1088/2040-8978/18/10/104003}, but sideband cooling in the mechanically degenerate regime has not been yet been investigated experimentally.

In this paper, we demonstrate optomechanical cooling of an optomechanical system with two nearly degenerate mechanical modes. We show cooling of the individual mechanical modes at low drive powers, and cooling of the hybridized mechanical modes in the regime of mechanical degeneracy. Approaching the strong coupling regime, we find these modes can be described as a bright mechanical mode with increasing optomechanical damping rate, and a dark mode with a damping rate approaching the intrinsic mechanical linewidth. While the occupation of both modes increases at high pump powers due to technical heating effects, we observe that the bright modes remain colder than the dark mode.

\begin{figure}
\includegraphics[width=\columnwidth]{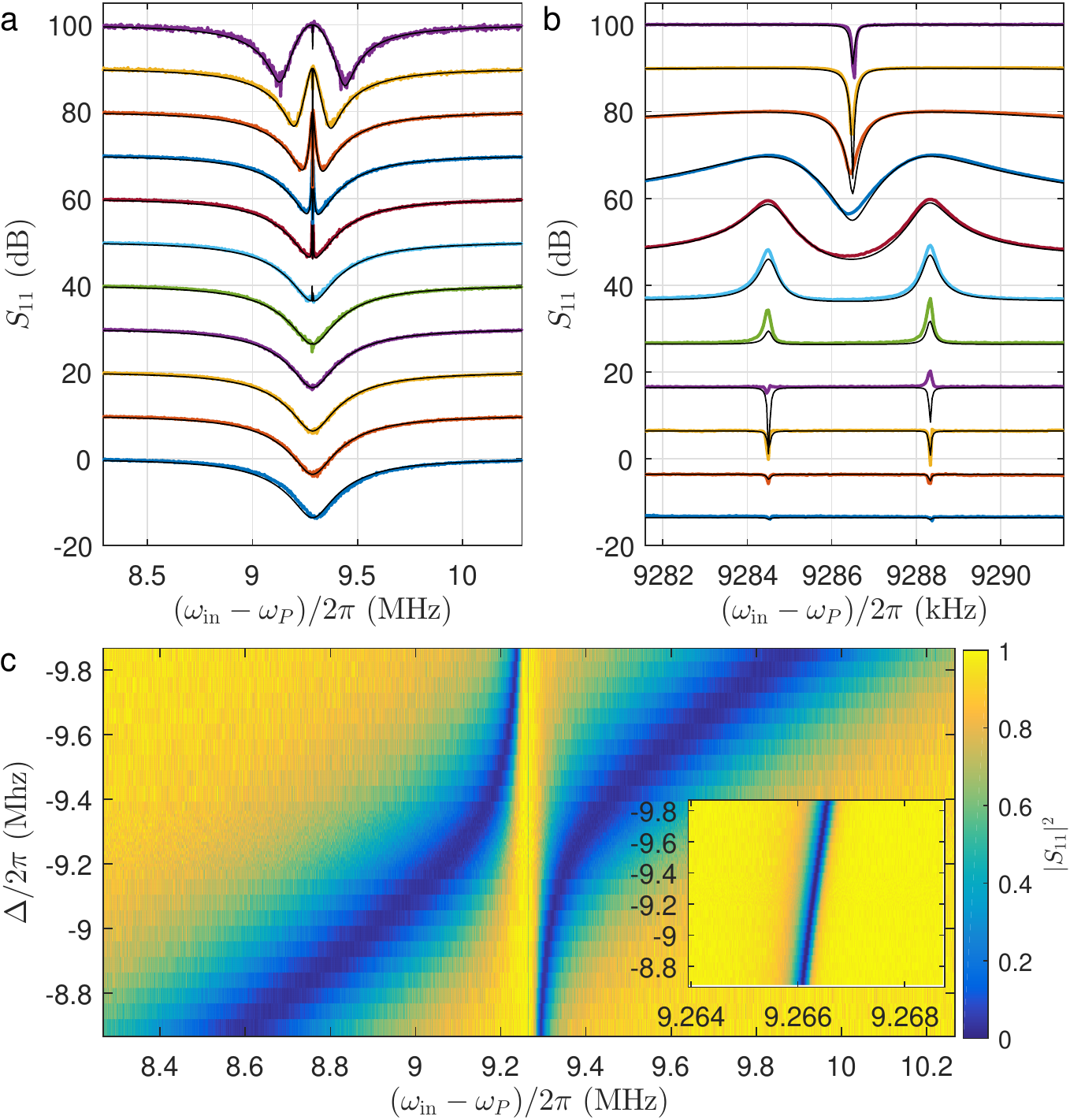}
\caption{\emph{Linear response.} (a) Cavity linear response for pump powers increasing in steps of 5 dB from bottom to top, corresponding to $G_1 = 2\pi\times357~$Hz (bottom curve) to $G_1 = 2\pi\times113~$kHz (top curve). Each curve is vertically offset by 10~dB. Black curves are a theory prediction (see text). At high pump powers, splitting of the spectrum into hybrid normal modes is visible. (b) Detail of panel a, showing anti-Stokes peaks of the individual mechanical modes at low power, and the formation of a single dark mode at high power. (c) Response as function of pump detuning for $G_1 \approx 2\pi\times94~$kHz. Inset: detail showing dark mode resonance.}
\label{fig:s11}
\end{figure}

Our experiments are performed in a microwave optomechanical system \cite{doi:10.1038/nature10261,doi:10.1038/s41586-018-0038-x}, where the effective microwave cavity is an aluminium superconducting circuit on a quartz substrate, resonant at $\omega_c = 2\pi\times 4.2~$GHz. The mechanical modes are two drum resonators which act in the circuit as compliant capacitances. Hence, the mechanical position affects the cavity frequency, realizing the optomechanical system of Eq.~\eqref{eq:hamiltonian}. We use the fundamental mechanical mode of each drum with frequency $\omega_{1,2} = 2\pi\times 9.2~$MHz. The two drums have nearly identical frequencies, separated by only $\omega_2 - \omega_1 = 2\pi\times 3.9~$kHz, and intrinsic energy decay rates of $\gamma_1 = 2\pi\times 105\pm8~$Hz and $\gamma_2 = 2\pi\times 71\pm19~$Hz. Hence, the modes are initially distinct, but the mechanically degenerate regime can be reached with moderate pumping strength. The total cavity linewidth $\kappa = \kappa_i + \kappa_e$ consists of the intrinsic decay rate $\kappa_i = 2\pi\times 268$~kHz and a coupling rate to the microwave feedline of $\kappa_e = 2\pi\times 410$~kHz. 

We measure the cavity in reflection as shown schematically in Fig.~\ref{fig:setup}, in a dry dilution refrigerator at a temperature of $T \approx 20~$mK. The input signal, consisting of the strong pump tone (frequency $\omega_P$) and optionally a weak probe tone (frequency $\omega_\text{in}$) from a vector network analyzer (VNA). The reflected signal is amplified at 4~K by a low-noise amplifier, and further amplified at room temperature before being recorded on the VNA or on a signal analyzer.

\begin{figure*}
\includegraphics[width=\textwidth]{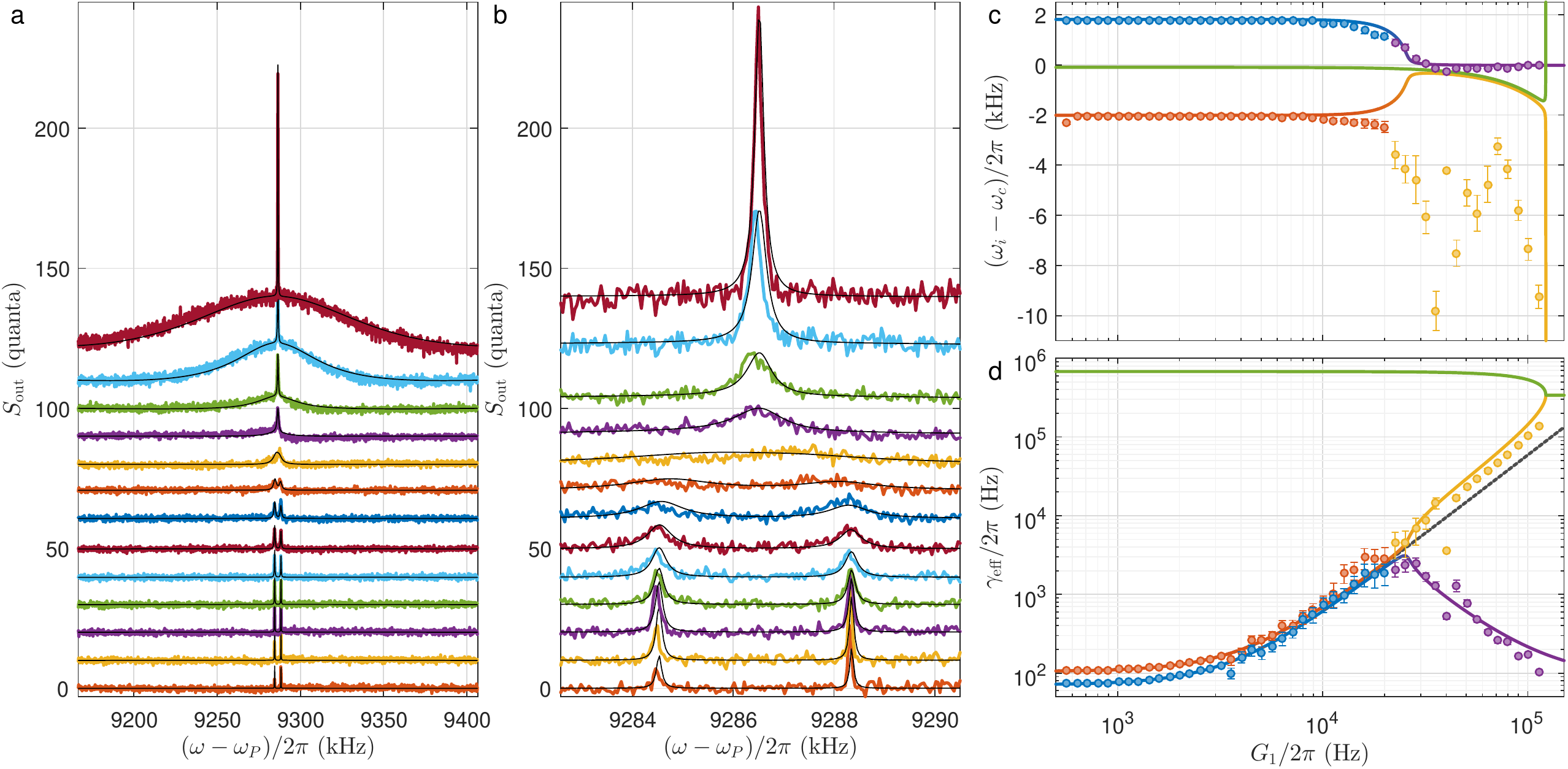}
\caption{\emph{Noise spectrum.} (a) Cavity output spectrum for varying pump power ranging from $G_1 = 1.6~$kHz (lower curve) to $G_1 = 101~$kHz (upper curve) in 3~dB steps. Black lines show a theory fit (see text). For each curve the background level is subtracted, and subsequently vertically offset by 10 units for clarity. (b) Detail of panel a. (c) Frequency and (d) linewidth of the eigenmodes. Circles correspond to Lorentzian fits to the data, solid lines show the predicted eigenmodes from the three-mode model. At small $G_1$ these correspond to the mechanical (red, blue) and cavity (green) modes, respectively; at large $G_1$ the mechanical modes mix to dark (purple) and bright (yellow) modes. Dotted line shows the expected linewidth for a system with a single mechanical oscillator. Error bars show statistical 95\% confidence levels.}\label{fig:sout}
\end{figure*}

We first study the linear cavity response in the presence of the strong pump. We tune the pump tone to the average of the red mechanical sidebands, $\omega_P = \omega_c - (\omega_1 + \omega_2)/2$, and measure the response $S_{11}$ with the VNA. Figure~\ref{fig:s11}a,b shows the measured response for varying pump powers. For small pump powers the mechanical modes remain uncoupled, and the anti-Stokes scattering process is visible as two peaks or dips near the cavity center. The precise shape of these features depends on interference between the anti-Stokes scattered pump tone and the reflected probe tone, commonly referred to as optomechanically induced transparency and related 
effects~\cite{doi:10.1126/science.1195596,doi:10.1038/nnano.2014.168}.

At large pump powers, the cavity mode splits into two broad bright modes, and the dark mode appears as a narrow dip in the cavity center. At the highest pump powers, we reach the onset of the strong coupling regime for the bright modes. In Fig.~\ref{fig:s11}c the response at high pump power is shown as function of pump detuning. 
The bright modes show an avoided crossing similar to that observed with a single mechanical mode~\cite{doi:10.1038/nature09898}, while the dark mode can be seen as a narrow feature nearly independent of detuning. An analytic expression for the cavity response was derived from Eq.~\eqref{eq:linearhamiltonian} in Ref.~\cite{doi:10.1038/ncomms1993}. As shown in Fig.~\ref{fig:s11}a,b, the measured data are very well described by this model. For the shown theory curves, only an overall scaling of $G_i^2$ with applied power was used as an adjustable parameter (see appendix).

Next, we study the output spectrum and optomechanical cooling. The anti-Stokes scattering process transduces the thermal mechanical motion to the cavity output spectrum, producing for a single oscillator a peak in the spectrum whose power (area under the peak) $\propto n_P n_i$, where $n_i$ is the phonon occupation of the mechanical mode $i$.  The measured data are shown in Figure~\ref{fig:sout}a,b. Here, the amplifier noise floor was subtracted from the measured spectrum. At low pump power, the standard sideband cooling spectrum is observed for both individual modes, with increasing linewidth as cooling power is increased. For $G_{1,2}\gtrsim20$~kHz, the individual modes overlap and the bright and dark modes are visible as a broad and narrow peak, respectively. 

The output spectrum due to sideband cooling of a single mechanical mode is well known, and provides information on the position of the mechanical oscillator. For low pump powers we can treat our oscillators as independent modes, and fit each peak to the standard Lorentzian form~\cite{doi:10.1038/nature08681}
\begin{equation}\label{eq:soutcool}
S_\text{out}(\delta) = \frac{4\kappa_e}{\kappa} n_I^T + \gamma_\text{opt} \frac{\kappa_e}{\kappa} \frac{\gamma_\text{eff}}{\delta^2 + \gamma_\text{eff}^2/4} \left(n_i - 2n_I^T\right),
\end{equation}
where $n_I^T$ is the thermal occupation of the cavity and $\delta = \omega - \omega_c - \delta_i$ is the frequency relative to the cavity. Here we introduced an additional detuning $\delta_i$ which incorporates a small difference between the pump frequency and the red sideband in our two-mode configuration, as well as any frequency shift of the modes due to the optical spring effect and a small power dependence of the cavity frequency. Equation~\eqref{eq:soutcool} is valid in the resolved sideband regime and for frequencies $\delta \ll \kappa$ close to the cavity center. 

In the regime of degenerate mechanical modes, $G_{1,2}\gtrsim20$~kHz, we cannot treat the mechanical oscillators independently. To model the complete three-mode system we write the quantum Langevin equations of motion (EOMs) corresponding to Equation~\eqref{eq:linearhamiltonian}, including the dissipation rates $\gamma_i$, $\kappa_i$, and $\kappa_e$. We use standard input-output theory to numerically predict the thermal output spectrum of the cavity. The thermal drive is characterized by effective bath occupation numbers $n_{m}^T$ for the mechanical modes, $n_I^T$ for internal cavity heating and $n_E^T$ for external cavity heating. As shown in figure~\ref{fig:sout}a,b, the predicted output spectrum is in excellent agreement with the data, using only the thermal occupation numbers as free parameters.

An interesting question is if the features of the output spectrum correspond directly to the hybridized eigenmodes of the three-mode system. To investigate this question, we fit the data at all pump powers with a sum of two Lorentzians, distinguishable either by frequency (low power) or line width (high power). Figures \ref{fig:sout}c and \ref{fig:sout}d show the fitted frequencies and line widths, respectively, compared to the predicted frequencies and damping rates of the eigenmodes obtained from our model. The fitted line widths and predicted damping rates are in very good agreement, indicating that there is indeed a correspondance between the bright and dark modes and the spectral features. The bright mode experiences an enhanced optomechanical damping rate compared to the bare mechanical oscillator. This is to be expected based on a general property that incoherently adding more coupled systems results in a coupling enhanced as the square root of the number of elements (here two). Conversely, the dark mode decouples from the optical system at high pump powers. The observed frequency of the bright mode shows a small deviation (smaller than the line width at these powers) from the predicted frequency, which may be due to an additional optical spring effect if the pump was not exactly at the sideband co-resonance condition.

\begin{figure}
\includegraphics[width=\columnwidth]{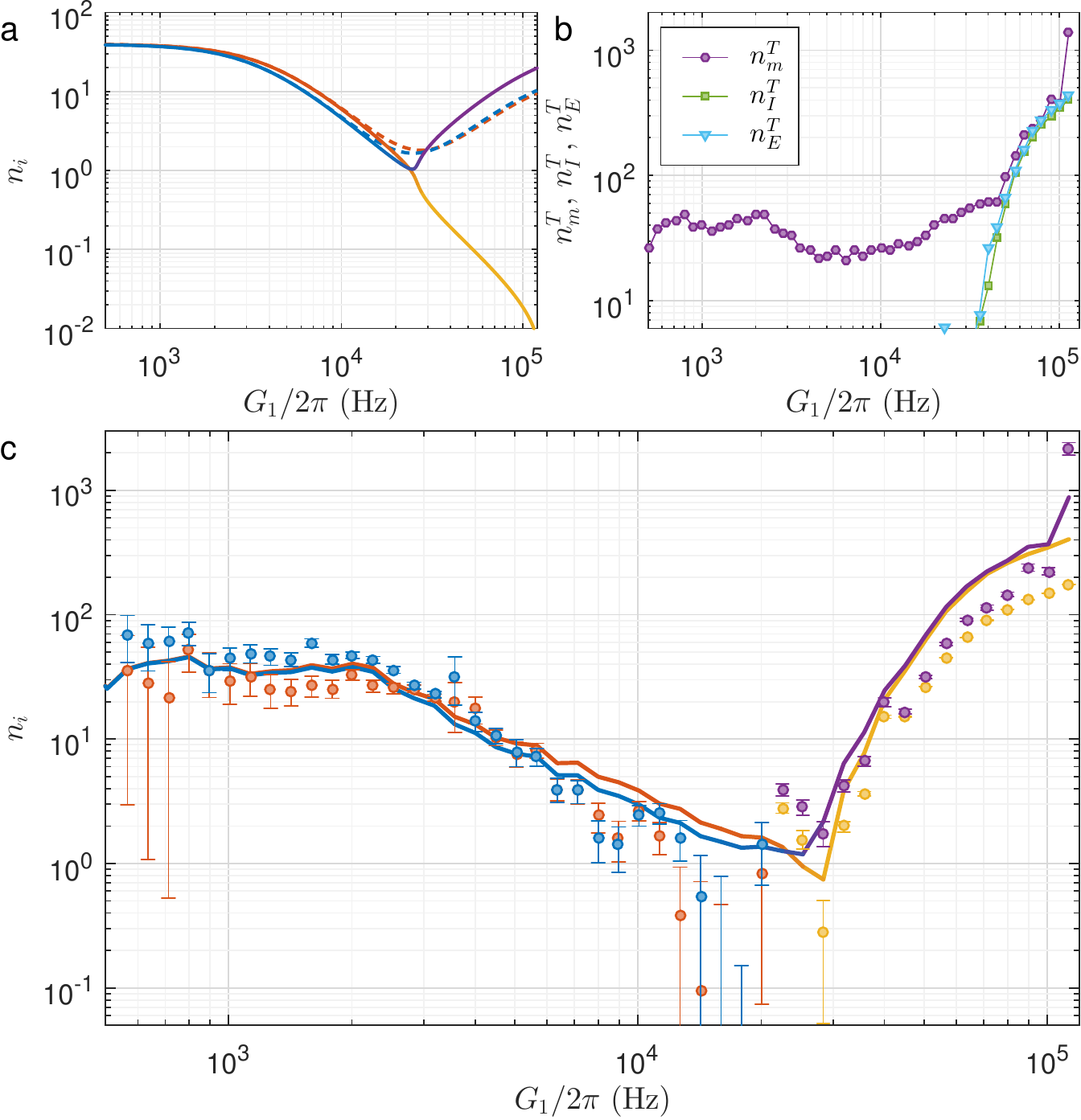}
\caption{\emph{Sideband cooling.} a) Theory prediction of the mode occupation of the bare mechanical modes (dashed lines) and the mechanic-like eigenmodes of the three-mode system (solid lines), in the absence of technical heating effects. Colors as in Fig.~\ref{fig:sout}b,c. b) Fitted effective environment temperatures in the experimental data, expressed in quanta. c) Occupation of the eigenmodes taking the fitted environment temperatures into account (solid lines). Circles show the temperatures inferred based on Lorentzian fits to the data according to Eq.~\eqref{eq:soutcool}. The data were normalized to correspond to the fridge temperature at low pumping power. Error bars show statistical 95\% confidence levels. Colors as in Fig.~\ref{fig:sout}b,c
}\label{fig:n}
\end{figure}

Finally, we discuss sideband cooling in the tree-mode system. We calculate the expected energy spectra of the bare mechanical modes $b_i$ in the presence of thermal drive and the pump tone using the three-mode model, and numerically integrate these spectra to obtain an effective occupation number of the mechanical modes. We first consider the case $n_{m}^T = 40$ and $n_I^T = n_E^T = 0$, corresponding to the fridge temperature. As shown by in Fig.~\ref{fig:n}a (dashed lines), both mechanical modes are expected to cool at low powers, but sideband cooling breaks down in the degenerate regime, in agreement with Ref.~\cite{doi:10.1038/ncomms1993}. However, we can alternatively calculate the spectra of the eigenmodes of the three-mode system. The predicted occupation numbers corresponding to the mechanic-like eigenmodes are shown as solid lines in Fig.~\ref{fig:n}a. In the mechanically degenerate regime, the dark mode re-thermalizes with the environment, but the bright mode can be cooled deep into the quantum regime. We note that in this degenerate regime the bright mode is still predominantly mechanical, with up to 20\% of the cavity mode mixing in at $G1 \approx 2\pi\times 100~$kHz.

To estimate the occupation of the eigenmodes in the experiment, we use the same calculation with the environment temperatures obtained from fitting the experimental data (Fig.~\ref{fig:sout}a). As shown in Fig.~\ref{fig:n}b, the thermal environment of all modes was found to heat up significantly in the experiment. This technical heating most likely originated from phase noise of the pump generator, which drives the system with an approximately flat spectrum across the narrow cavity line width similar to thermal excitation. The estimated mode ocupations are shown in Fig.~\ref{fig:n}c. The technical heating causes significant heating of both mechanic-like eigenmodes. Nevertheless, the bright mode remains colder overall.

For comparison, we also show the mode occupation numbers obtained from fitting a sum of two Lorentzians according to Eq.~\eqref{eq:soutcool} to the data. The results are in in fairly good agreement with the full three-mode model, even at high power corresponding to hybrid bright and dark modes. In the latter case the physical interpretation of Eq.~\eqref{eq:soutcool} is not clear since it does not take external cavity noise into account. However, by comparison to the full model we verified that in the absence of technical cavity heating Eq.~\eqref{eq:soutcool} is in good agreement with the three-mode model.

In conclusion, we have experimentally demonstrated sideband cooling in a near-degenerate multimode optomechanical system. As has been predicted theoretically, cooling of the individual mechanical oscillators is suppressed. However, we find that the optically bright hybrid mode can be cooled differentially with respect to the dark mode. Furthermore, the bright mode has enhanced optomechanical coupling, while the dark mode decouples from the system. We show theoretically that in the absence of technical heating effects the bright mode could be cooled deeply into the quantum ground state. These results are relevant to high-overtone optomechanics, where multiple closely spaced mechanical modes could be hybridized to realize a single, strongly coupled and cooled bright supermode. 

\appendix*
\section{Calibrations and data analysis}
We first calibrate $\omega_c$, $\kappa_i$ and $\kappa_e$ from the cavity linear response with no pump tone present. From measurements with a weak pump tone, we measure the mechanical frequencies and make an initial estimate on $\gamma_i$. With these calibrations in place, the only free parameters left to describe the pumped cavity response (Fig.~\ref{fig:s11}) are $G_1$ and $G_2$. Since we do not accurately know the attenuation of the input cables, we cannot infer the values of $g_i$ directly, but the spectra are sensitive to their ratio and we find  that both mechanical modes experience nearly equal coupling, $g_2/g_1 = 0.95 \pm 0.05$. Finally, we use the ratio between $G_1^2$ and the power setting of the generator as a single free parameter, which we fit with an accuracy of $\sim 0.1~$dB. The data shown in Fig.~\ref{fig:s11}c was measured in a separate cool-down of the same sample, resulting in slightly different parameters that were calibrated separately.

It is generally difficult to accurately fit the internal mechanical linewidths $\gamma_{1,2}$, since at the required weak pump powers the signal-to-noise ratio is limited. We obtain the best estimate for $\gamma_{1,2}$ in the regime of intermediate pump powers, where the modes are well separated but experience significant cooling ($\gamma_\text{opt}/2\pi \sim 20\ldots100~$Hz). Using the previously described calibration for $G_{1,2}$, we subtract the extra dissipation $\gamma_\text{opt}$ from the measured linewidths, and obtain the quoted estimates for $\gamma_{1,2}$.

At high pump powers $G_{1,2}\gtrsim 2\pi\times50~$kHz we observe saturation of the low-noise HEMT amplifier. We measure the saturation, up to $13~$dB at the highest powers used in the experiments, from the off-resonant linear response and use this measurement to correct both the linear response and output spectrum data. We note that while the strong pump tone affects the gain of the amplifier at all frequencies, we observe no nonlinearity at the cavity frequency, as witnessed by the agreement between measurement and theory in Fig.~\ref{fig:s11}.

\begin{acknowledgments}
This work was supported by the Academy of Finland (contract 250280, CoE LTQ, 275245, 308290) and by the European Research Council under grant 615755-CAVITYQPD and under the European Union's H2020 programme/ERC Grant Agreement 681476-QOM3D. We acknowledge funding from the European Union's Horizon 2020 research and innovation program under grant agreement number 732894 (FETPRO HOT). The work benefited from the facilities at the Micronova Nanofabrication Center and at the Low Temperature Laboratory infrastructure.
\end{acknowledgments}

\end{document}